\documentstyle[mprocl,psfig]{article}

\def\Journal#1#2#3#4{{#1} {\bf #2}, #3 (#4)}

\def\CQG{\em Class. Quantum Grav.}

\def\PRD{{\em Phys. Rev.} D}

\def\be{\begin{equation}}
\def\ee{\end{equation}}
\def\bea{\begin{eqnarray}}
\def\eea{\end{eqnarray}}
\def\bhds{black hole-de~Sitter }
\def\sds{Schwarzschild-de~Sitter }
\def\rnds{Reissner-Nordstr\"{o}m-de Sitter }
\def\etal{\it et al. }
\begin{document}

\title{SOME COSMOLOGICAL TAILS OF COLLAPSE}

\author{C.M. CHAMBERS}

\address{Department of Physics, Montana State University,
         Bozeman, MT 59717-3840, USA}

\author{P.R. BRADY}

\address{Theoretical Astrophysics, Mail-Code 130-33,
         California Institute of Technology, Pasadena,
	 CA 91125, USA}

\author{W. KRIVAN~\footnote{Also, Institut f\"{u}r 
        Astronomie und Astrophysik, Universit\"{a}t 
	T\"{u}bingen, D-72076 T\"{u}bingen,
	Germany.}, P. LAGUNA}

\address{Department of Astronomy \& Astrophysics and
         Center for Gravitational Physics \& Geometry,
	 Pennsylvania State University, University Park,
	 PA 16802, USA}
\maketitle

\abstracts{We summarize the results of an investigation
           into the late time behavior of massless
	   scalar fields propagating on spherically
	   symmetric black hole spacetimes with a 
	   non-zero cosmological constant. The 
	   compatibility of these results with
	   `minimal requirement' of Brady and 
	   Poisson~\cite{cmc-b10} is commented upon. } 

%
%

\section{Introduction}
Black holes immersed in de~Sitter space are known to provide
reasonable counter-examples to the strong cosmic censorship
hypothesis -- Analyses have shown that the Cauchy horizon is
stable provided the surface gravity of the cosmological event horizon
is greater than that of the inner horizon.\cite{cmc-b10} The key
ingredient leading to stability is the rapid decay of the {\it
radiative tails} of gravitational collapse at late times.  For black
holes in asymptotically flat spacetimes, the form of these tails is
well known,\cite{cmc-b20} and is largely responsible for the weakness
of the mass-inflation singularity inside charged black
holes.\cite{cmc-b30} However, for \bhds spacetimes, the form of the
radiative tails was previously unknown, although the
physically reasonable requirement of a finite but non-vanishing
flux at the cosmological event horizon suggests the tails should decay
exponentially, with a folding time given by the surface gravity
$\kappa_{1}$ of the cosmological event horizon.  Given
the importance of tails to the current picture of 
Cauchy horizon stability in \bhds spacetimes and their key role as 
initial data in numerical studies of the black-hole interior, 
it seemed imperative to confirm their precise form.

\section{Results}

\begin{figure}[ht]
	\centerline{\hbox{
	\psfig{figure=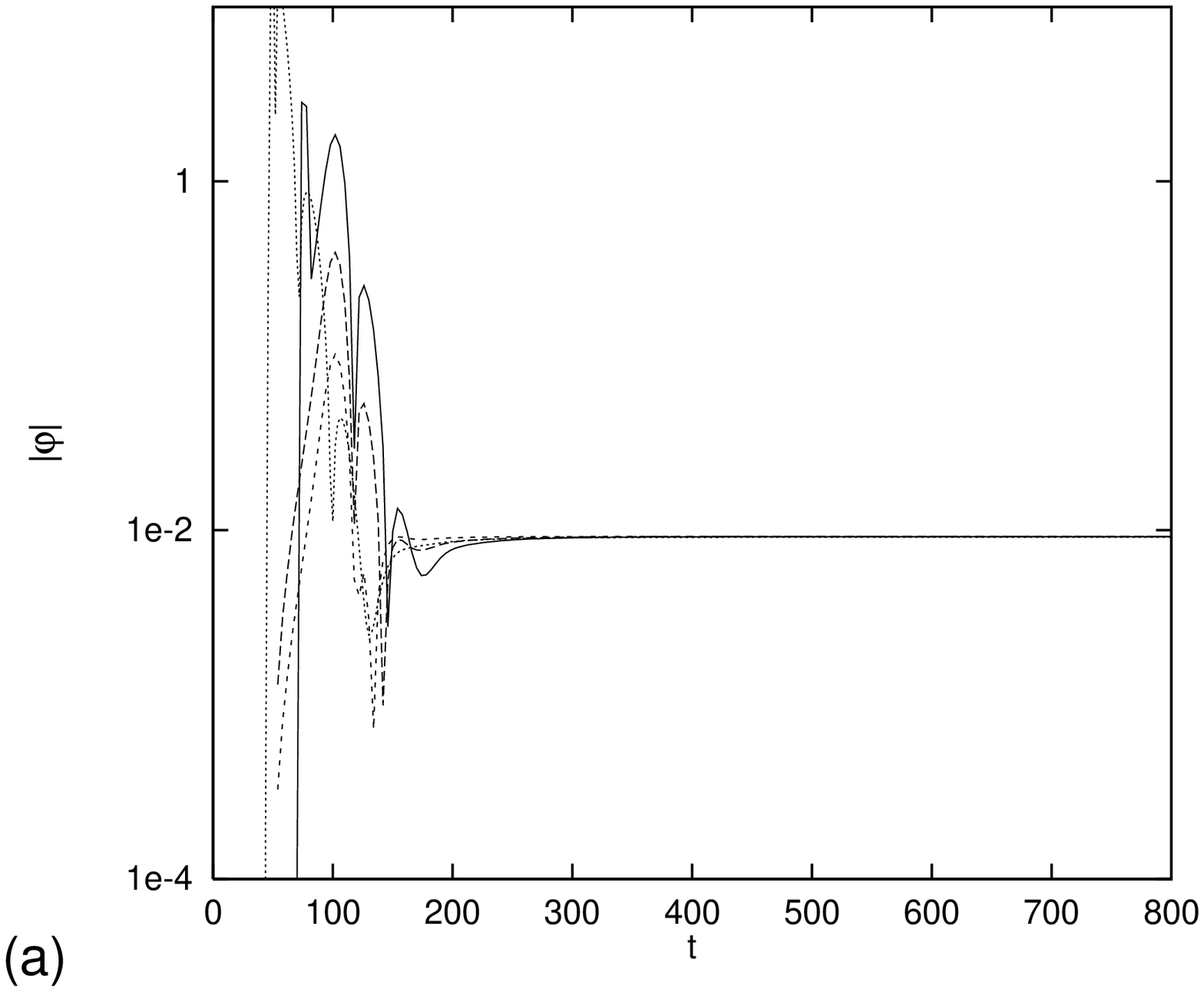,height=2.0in}
	\psfig{figure=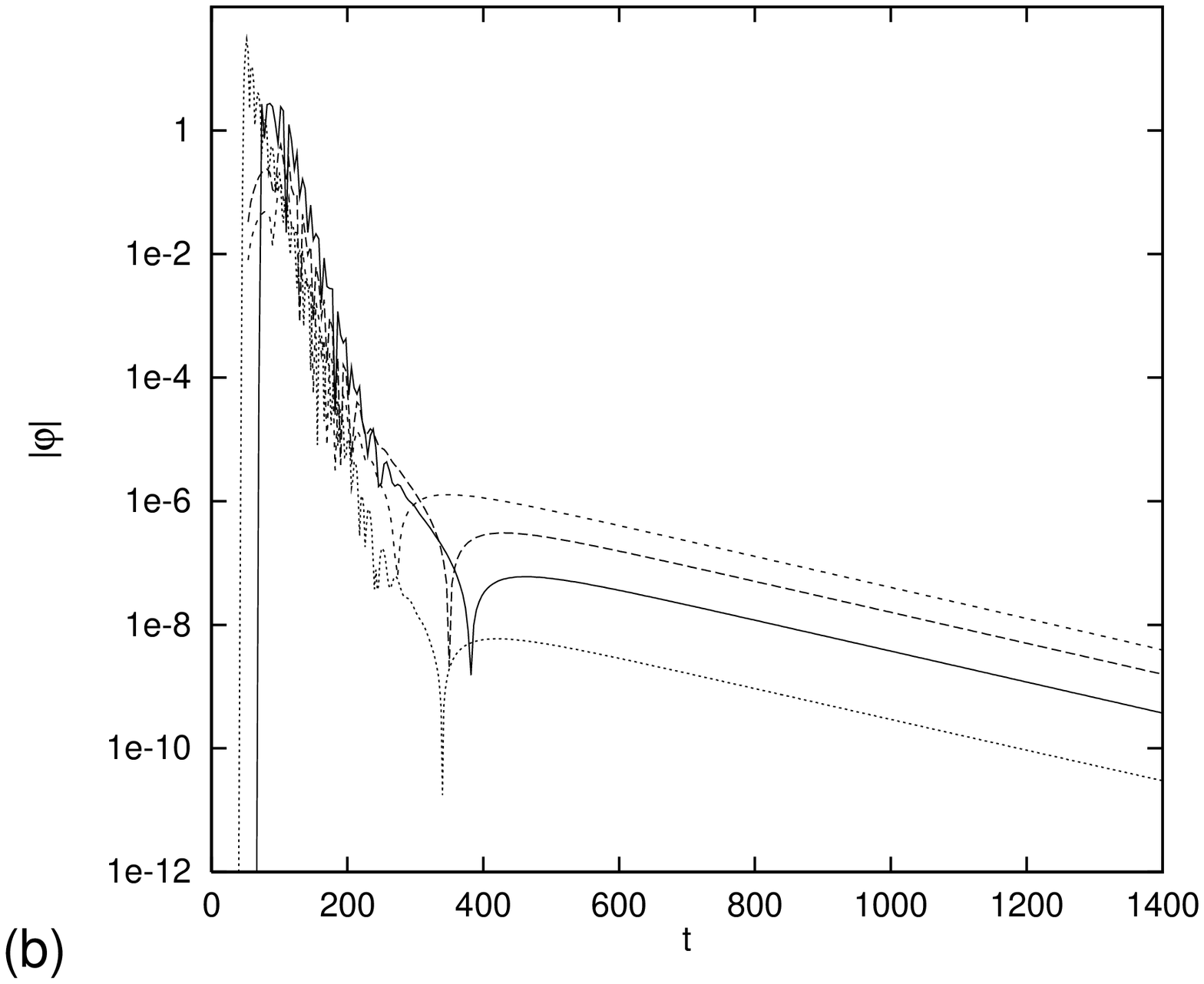,height=2.0in}
	}}
	\caption{Plots of $|\varphi|$ versus t for a
	\rnds black hole. 
	(a) Field behavior for an $\ell = 0$ mode.
	(b) Field behavior for an $\ell = 1$ mode. Details
	are given in the text. \label{cmc-f10}}
	\end{figure}

We have studied the evolution of massless, minimally coupled scalar
fields propagating on spherically symmetric spacetimes with a non-zero
cosmological constant numerically,\cite{cmc-b40} using two standard
methods:~\cite{cmc-b50} (i) A linear test-field analysis, 
with the field propagating on the fixed backgrounds of \sds and \rnds
spacetimes, and (ii) A non-linear method, where the field
is coupled to a general spherically symmetric spacetime
though the Einstein-Klein-Gordon field equations. In general we 
have found that an $\ell$-pole mode of the field decays, at late 
times, like 
\be
\varphi_{\ell} \sim e^{-\ell \kappa_{1} t} \ \ \ \ \ell > 0 \; ,
\label{cmc-e10} 
\ee 
where $t$ is the time coordinate.  For $\ell = 0$,
we find that rather than the decay, the field approaches a constant
value 
\be 
\varphi_{\ell = 0} \sim \varphi_{0} + \varphi_{1}(r) e^{-2
\kappa_{1} t} \ \ \ \ {\rm as} \ \ \ \ t \rightarrow \infty \; ,
\label{cmc-e20} 
\ee 
where $\varphi_{0}$ scales like
$\Lambda$.\cite{cmc-b40} Figure~\ref{cmc-f10} shows plots of
$|\varphi|$ versus $t$ for a \rnds back hole with charge ($e$) of $0.5$,
mass ($M$) of $1.0$ and a cosmological constant ($\Lambda$) of
$10^{-4}$, the results being representative of a more general choice
of $(e,M,\Lambda)$.  The field behavior is shown along four different
surfaces:~\footnote{Since the field behavior is identical along all
four surfaces we do not distinguish them here. More detail can be
found in the work of Brady \etal\cite{cmc-b40}} (a) The black hole
event horizon.  (b) The cosmological event horizon.  (c) Future
timelike infinity -- approached along two surfaces of constant radius.
Figure~\ref{cmc-f10}~(a) shows the behavior of the $\ell=0$ mode of
the scalar field. At early times $(0<t<200)$ the field is dominated by
quasinormal ringing, while at late times $(t>200)$ this behavior is
swamped by the radiative tails. It is clear that the field
$\varphi_{\ell=0}$ approaches the same constant value
[Eq.~\ref{cmc-e20}], along each surface, at late times.  A numerical
investigation of the dependence of $\varphi_{0}$ on $\Lambda$
provides the scaling relation $\varphi_{0} \propto
\Lambda^{0.9994}$. An analytic calculation shows that $\varphi_{0}$
varies like $\Lambda$, confirming the numerical result.  Examining
$\varphi_{,t}$ has allowed us to further conclude that the exponent in
Eq.~\ref{cmc-e20} is indeed $\kappa_{1}$ to within
$10\%$. Figure~\ref{cmc-f10}~(b) shows the behavior of the $\ell=1$
mode of the scalar field. Again, at early times $(0<t<400)$, the field
decay is dominated by quasinormal ringing. At late times though
$(t>400)$, the behavior is swamped by an exponential decay, with the
field falling off as $\exp(-k t)$. Numerically we have ascertained
that $k \approx \kappa_{1}$ to within $2 \%$. An examination of higher
modes has allowed us to conclude that in general an $\ell$-pole mode
decays according to Eq.~\ref{cmc-e10}.
%
%
\section{Conclusions}
Our numerical results demonstrate that at late times the scalar field can be  
expanded as 
	\[
	  \varphi(t,r) = \varphi_{0} + \sum_{n=1}^{\infty}
	  \varphi_{n}(r) e^{-n \kappa_{1} t} \ \ \ \ 
	  {\rm as} \ \ \ \ t \rightarrow \infty \; .
        \]
The minimal requirement of Brady and Poisson, requires the field 
to have the late time form~\cite{cmc-b60}
	\[
	  \varphi (t,r) \sim a(r) + b(r) e^{-\kappa_{1} t} \; .
	\]
Making the identifications $a(r) = \varphi_{0}$ and $b(r) = 
\varphi_{1}$ demonstrates the correctness of the Brady-Poisson
model.\cite{cmc-b10} It is interesting to note that it is the 
$\ell = 1$ behavior that provides the required form for the
radiative tails. The contribution from the other $\ell$ modes
leads to a vanishing flux at the Cauchy horizon. 

\section*{Acknowledgments}

CMC is a Fellow of The Royal Commission for the Exhibition
of 1851 and gratefully acknowledges their financial support.
Travel to Israel for CMC was supported by NSF Grant Nos. 
PHY-9722529, to North Carolina State University, and 
PHY-9511794 to Montana State University. 
PRB is supported by a PMA Division Fellowship at Caltech and
NSF Grant No. AST-9417371. The work of WK was supported by
the Deutscher Akademischer Austauschdienst (DAAD). PL is
supported by the Binary Black Hole Grand Challenge Alliance,
NSF PHY/ASC 9318152 (ARA supplemented) and by NSF grants PHY
96-01413, 93-57219 (NYI).


\section*{References}
\begin{thebibliography}{99}
\bibitem{cmc-b10}
P.R. Brady and E. Poisson,
\Journal{\CQG}{9}{121}{1992}.
\bibitem{cmc-b20}
R.H. Price,
\Journal{\PRD}{5}{2419}{1972}; \Journal{\PRD}{5}{2439}{1972}.
\bibitem{cmc-b30}
E. Poisson and W. Israel,
\Journal{\PRD}{41}{1796}{1990}.
\bibitem{cmc-b40}
P.R. Brady, C.M. Chambers, W. Krivan and P. Laguna,
\Journal{\PRD}{55}{5558}{1995}.
\bibitem{cmc-b50}
C. Gundlach, R.H. Price and J. Pullin,
\Journal{\PRD}{49}{883}{1994};
\Journal{\PRD}{49}{890}{1994}.
\bibitem{cmc-b60}
C.M. Chambers, {\em The Cauchy Horizon In Black Hole-de~Sitter
Spacetimes}, gr-qc9709025.
\end{thebibliography}
\end{document}